\documentclass[aps,prb,twocolumn,superscriptaddress,floatfix]{revtex4}
\usepackage{color}
\usepackage{graphicx}
\usepackage{epsfig}
\usepackage{textcmds}
\bibstyle{apsrev.bib}
\usepackage{amsmath,amssymb}
\usepackage{graphicx}

\begin{document}
\input epsf.sty

\title{Interface mapping in two-dimensional random lattice models}

\author{M. Karsai}
\affiliation{
Institute of Theoretical Physics,
Szeged University, H-6720 Szeged, Hungary}
\affiliation{Institut N\'eel-MCBT
 CNRS 
\thanks{U.P.R. 5001 du CNRS, Laboratoire conventionn\'e
avec l'Universit\'e Joseph Fourier}, B. P. 166, F-38042 Grenoble,
France}
\affiliation{Department of Biomedical Engineering and Computational Science, Aalto 
University
Centre of Excellence in Computational Complex System Research
P.O. Box 12200 FI-00076 Aalto, Finland}

\author{J-Ch. Angl\`es d'Auriac}
\affiliation{Institut N\'eel-MCBT
 CNRS 
\thanks{U.P.R. 5001 du CNRS, Laboratoire conventionn\'e
avec l'Universit\'e Joseph Fourier}, B. P. 166, F-38042 Grenoble,
France}
\affiliation{Laboratoire de Physique Subatomique et de Cosmologie 53, av. des Martyrs F-30026 Grenoble,France}
\author{F. Igl\'oi}
\affiliation{
Research Institute for Solid State Physics and Optics,
H-1525 Budapest, P.O.Box 49, Hungary}
\affiliation{
Institute of Theoretical Physics,
Szeged University, H-6720 Szeged, Hungary}

\date{\today}

\begin{abstract}
We consider two disordered lattice models on the square lattice: on
the medial lattice the random field Ising model at $T=0$ and on the
direct lattice the random bond Potts model in the large-$q$ limit at
its transition point. The interface properties of the two models are
known to be related by a mapping which is valid in the continuum
approximation.  Here we consider finite random samples with the same
form of disorder for both models and calculate the respective
equilibrium states exactly by combinatorial optimization
algorithms. We study the evolution of the interfaces with the strength
of disorder and analyse and compare the interfaces of the two models in
finite lattices.

\end{abstract}

\maketitle

\newcommand{\bc}{\begin{center}}
\newcommand{\ec}{\end{center}}
\newcommand{\be}{\begin{equation}}
\newcommand{\ee}{\end{equation}}
\newcommand{\beqn}{\begin{eqnarray}}
\newcommand{\eeqn}{\end{eqnarray}}

\section{Introduction}
\label{sec:Intr}
The presence of different type of quenched disorder could influence
the cooperative behaviour of many-body systems in different
measure. In this respect dilution or random ferromagnetic bonds, which
are often called as $T_c$-disorder, have a relatively weak effect. The
ferromagnetic order in the pure system, which is present for $T<T_c$,
is not destroyed by a small amount of $T_c$-disorder. Dilution has a
dramatic effect only in the vicinity of a phase transition point,
$|T-T_c|/T_c \ll 1$.  If this transition is of second order even its
universality class can be changed for any small amount of
disorder\cite{harris}.

A disorder perturbation with a stronger effect is represented by
random fields\cite{natt}. As an example we consider here the random
field Ising model (RFIM) in which the fields are taken from a
distribution with zero mean and variance $h_0^2$.  In dimensions,
$d>2$, the phase transition is governed by a zero-temperature fixed
point, the ground state of the system being disordered for
$h_0>h_{0,c}(d)>0$, and there is magnetic long-range-order for
$h_0<h_{0,c}(d)$. In two-dimensions we have $h_{0,c}(2)=0$, thus the
magnetic order in the ground state is destroyed by any finite random
field\cite{imryma,aizenmanwehr}. The ground state of the RFIM is
conveniently characterized by geometric (or Ising)
clusters\cite{droplet}, which are formed by adjacent spins being in
the same orientation. In $2d$ these clusters are homogeneous only up
to a finite characteristic length, $\ell_g$, which is called the
breaking-up length of geometrical clusters\cite{lb,seppala,kornyei}.
In a coarse-grained picture the ground state consists of randomly
oriented domains of typical size, $\ell_g$, and as a consequence there
is no long-range-order in the system.  For weak disorder, $h_0 \ll J$,
$J=1$ being the nearest neighbour coupling constant, the breaking-up
length is large, and for $h_0 \to 0$ it diverges as\cite{lb}:
\be
 \ell_g(h_0) \sim \exp(A h_0^{-2})\;,
\label{breaking}
\ee
with a disorder dependent constant, $A=O(1)$.

As far as first-order transitions are concerned the effect of bond
disorder is also strong and the physical phenomena is analogous to
that of the RFIM discussed before\cite{Cardy99}.  As a prototypical
model having a first-order phase transition we consider the $q$-state
Potts model\cite{Wu} in which the number of states is sufficiently
large, $q>q_c(d)$.  In the pure system having homogeneous couplings
($J=J_0$) at the first-order transition point, $T=T_c$, the ordered
and disordered phases coexist, the latent heat is finite and there is
a finite jump of the order-parameter.  Introducing random
ferromagnetic couplings, such that the fluctuating part,
$\kappa=\dfrac{J-J_0}{J_0}$, has zero mean and variance $\kappa_0^2$
the latent heat and the jump of the magnetization is generally reduced
with increasing $\kappa_0$.  In dimensions $d>2$ for weak disorder,
$\kappa_{0,c}(d)>\kappa_0>0$, the transition stays first order,
whereas for strong disorder, $\kappa_0>\kappa_{0,c}(d)$, the latent
heat is vanishing and the transition softens to second
order\cite{imrywortis,aizenmanwehr}. The critical exponents of this
transition are independent of the strength of disorder, but they
generally depend on the value of
$q$\cite{pottsmc,pottstm,olson99,ai03,long2d}. In the borderline case,
$\kappa_0=\kappa_{0,c}(d)$, there is a tricritical point with a new
class of tricritical exponents, which are expected to be independent
of the value of $q$ and the actual fixed point is at $q \to
\infty$\cite{mai05}.  In $2d$ the limiting value is
$\kappa_{0,c}(2)=0$ and the phase transition is softened to second
order by any weak bond disorder\cite{imrywortis,aizenmanwehr}.

To characterize the state of the system we consider the
high-temperature series expansion\cite{kasteleyn}, which in the
large-$q$ limit is dominated by a single diagram. In the pure case
this diagram is either the fully connected graph, which happens in the
ordered phase, $T<T_c$, or the fully disconnected graph, which is the
case in the disordered phase, $T>T_c$.  At the phase transition point
these two basic graphs coexist. In the disordered model the dominant
diagram at $T=T_c$ contains both connected and disconnected parts. In
$2d$ the optimal diagram is homogeneous only up to a characteristic
finite length, $\ell_d$, which is the breaking-up length of the
optimal diagram. For small disorder the breaking-up length is large
and in the limit $\kappa_0 \to 0$ it is divergent\cite{mai05}: \be
\ell_d(\kappa_0) \sim \exp(B \kappa_0^{-2})\;,
\label{breaking1}
\ee
which is in the same form, as for the RFIM in Eq.(\ref{breaking}).  In
the thermodynamic limit the optimal diagram is a random composition of
the two basic graphs of typical size, $\ell_d$, therefore there is no
phase coexistence in $2d$ and the transition is of second order.

As described above the RFIM and the random bond Potts model (RBPM)
have analogous physical properties.  This analogy was first noticed by
Cardy and Jacobsen~\cite{pottstm} (CJ) and they have obtained an exact
mapping, which is valid in the limit of $d \to 2$ and $q \to \infty$,
when the interface Hamiltonian of the two problems have the same type
of solid-on-solid (SOS) description. From this mapping follows that
the energy exponents of the tricritical RBPM are equivalent to the
magnetization exponents of the critical RFIM.  Furthermore, analyzing
the renormalization group (RG) flow it was conjectured that the above
mapping stays valid for $d > 2$, in particular at $d=3$. This
conjecture has been numerically tested~\cite{mai05} for the $3d$ RBPM
in the large-$q$ limit.

The CJ-mapping has consequences at $d=2$, as far as interface
properties of the two models are concerned at large scales. This has
already been mentioned with respect to the same form of the
breaking-up lengths in Eqs.(\ref{breaking}) and (\ref{breaking1}). In
this context one may ask the question how this mapping works in finite
scales? In this paper we are going to perform a systematic study and
consider both models with exactly the same form of disorder. For this
we put the Potts model on the square lattice, whereas the Ising model
on the medial lattice. In this way couplings of the Potts model and
fields of the Ising model are on the same position. We study random
samples with fixed boundary conditions, which prefer the opposite
orientation for the RFIM and the different phases (basic graphs) for
the RBPM at the two halves of the lattice.  Using combinatorial
optimization algorithm\cite{heiko,aips02} we calculate exactly the
ground state of the RFIM and the optimal diagram of the RBPM and study
their evolution with increasing the strength of disorder. We also
compare the diagrams obtained in the two problems and study their
fitting.

The structure of the rest of the paper is the following. The RFIM and
the RBPM are defined in Sec.\ref{sec:Model} together with the
asymptotic mapping. Evolution of the diagrams with the strength of
disorder are presented in Sec.\ref{sec:Evol} whereas a comparison of
the diagrams of the two models are shown in Sec.\ref{sec:Comp}. Our
paper is closed by a discussion in Sec.\ref{sec:Disc}.

\section{Models and asymptotic mapping}
\label{sec:Model}
\begin{figure}[h]
\begin{center}
\includegraphics[width=7.cm,angle=0]{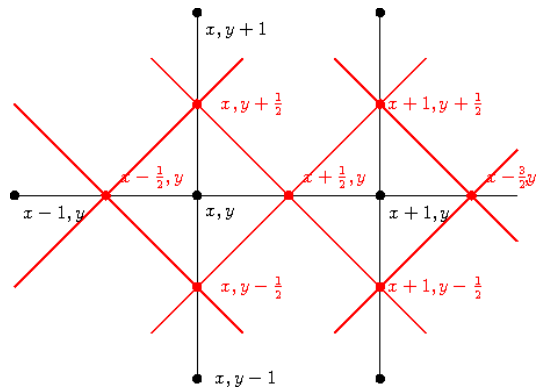}
\end{center}
\caption{ The square lattice (black) with the Potts spins and its
  medial lattice (red) with the Ising spins.
\label{fig1}
}   
\end{figure}

We consider a square lattice the sites of which are at $(x,y)$ with $1
\le x \le L$ and $-(L-1)/2 \le y \le (L-1)/2$ and both $x$ and $y$
being integer. In the corresponding medial lattice the sites are
placed to the middle of the links and have coordinates $(u,v)$ as
$(x,y + 1/2)$ and $(x + 1/2,y)$, see Fig.\ref{fig1}.


\subsection{Random field Ising model}

Spins of the random field Ising model, $\sigma_{u,v}=\pm 1$,
are put on the vertices of the medial lattice. The Hamiltonian of the
model is given by:
\be
{\cal H_I}=-\sum_{u,v} \sigma_{u,v} \sigma_{u+1/2,v \pm 1/2} -\sum_{u,v} h(u,v) \sigma_{u,v}\;.
\label{RFIM}
\ee
The external field, $h(u,v)$, is a random number which is parametrized as:
\be
h(u,v)=h_0 \epsilon(u,v)\;,
\label{h0}
\ee
where the distribution of $\epsilon(u,v)$ has zero mean and variance unity.
We apply fixed spin boundary conditions: the Ising spins at the upper part of the boundary ($v>0$) are fixed to $+1$,
and at the lower part of the boundary ($v<0$) they are fixed to $-1$. This model is studied at $T=0$. 
For a given realization of the disorder we calculate the ground state exactly by a combinatorial optimization
algorithm. 

\subsubsection{SOS approximation}

For weak disorder, such that $\ell_g(h_0)>L$, the ground state is well approximated by two clusters with $+1$ and $-1$
spins, respectively, and having an interface in between. Furthermore we use the solid-on solid (SOS) approximation, when the
height of the interface at position $u$ is given by a unique function, $\zeta(u)$. In this case the position of the interface
is obtained by minimizing the SOS energy functional\cite{pottstm}:
\be
E(\zeta)=- h_0 \sum_{{u}}\left(\sum_{v<\zeta(u)}-\sum_{v>\zeta(u)}\right)\epsilon(u,v)\;,
\label{SOS1}
\ee
with the constrain that the length of the interface, {\it measured in the medial lattice} is constant.

\subsection{Random bond Potts model}

The random bond Potts model is put on the vertices of the original square lattice and defined by the Hamiltonian:
\beqn
{\cal H_P}=-\sum_{x,y} &[&J(x,y+1/2)\delta(s_{x,y},s_{x,y+1})\\ \nonumber
&+&J(x+1/2,y) \delta(s_{x,y},s_{x+1,y})]\;.
\label{RBPM}
\eeqn
Here $s_{x,y}=1,2,\dots,q$ is a Potts spin variable at site $(x,y)$
and the couplings, $J(x,y + 1/2) \equiv J(u,v) >0$ and $J(x + 1/2,y) \equiv J(u,v) >0$,
are independent and identically distributed random numbers.
Introducing the reduced temperature, $T'=T \ln q$,
and similarly, $\beta'=1/T'=\beta/\ln q$, the partition function in the random cluster representation is given by\cite{kasteleyn}:
\begin{equation}
Z =\sum_{G\subseteq E}q^{c(G)}\prod_{(u,v)\in G}\left[q^{K(u,v)}-1\right]
\label{eq:kasfor}
\end{equation}
with $K(u,v)=\beta' J(u,v)$.
Here the sum runs over all subset of bonds, $G\subseteq E$ and $c(G)$ stands for the number of connected components of $G$.
In the following we consider the large-$q$ limit, where $q^{K(u,v)} \gg 1$, and the partition function can be written as
\begin{equation}
Z=\sum_{G\subseteq E}q^{\phi(G)},\quad \phi(G)=c(G) + \sum_{(u,v)\in G} K(u,v) \label{eq:kasfor1}
\end{equation}
which is dominated by the largest term, $\phi^*=\max_G \phi(G)$, so that\cite{JRI01,ai03}
\begin{equation}
Z = q^{\phi^*}+{\rm subleading ~ terms}.\label{eq:opti}
\end{equation}
For random couplings with $0 < K(u,v) < 1$ we use the form,
$K(u,v)=K(1+\kappa(u,v))$ and the $\kappa(u,v)$ random numbers are
taken from a symmetric distribution: $P(\kappa)=P(-\kappa)$, with a
variance $\kappa_0^2$, thus $\kappa(u,v)=\kappa_0 \epsilon(u,v)$. For
this type of distribution the phase transition point of the system
follows from self-duality and given by\cite{dom_kinz}:
\begin{equation}
K_c=\frac{1}{2}.
\end{equation}
Thus at the critical point the random couplings are parametrized as:
\be
\beta' J(u,v)=K(u,v)=\dfrac{1}{2}\left[ 1+\kappa_0 \epsilon(u,v)\right]\;
\label{kappa0}
\ee
and the strength of disorder is measured by $\kappa_0$.  In our
problem we use such type of boundary conditions, that the boundary
couplings at the upper part of the lattice $(v>0)$ are strong
$\kappa^{+}=1$, thus promoting an ordered phase, whereas at the lower
part of the lattice $(v<0)$ these are weak, $\kappa^{-}=0$, thus
favouring the disordered phase. For a given realization of the
disorder we calculate the ground state exactly by a combinatorial
optimization algorithm\cite{aips02}.

\subsubsection{SOS approximation}

In the following we consider the model at the critical point in the
weak disorder limit, when $\ell_d(\kappa_0)>L$.  In this case we use
the approximation that the optimal diagram consists of a fully
connected part, denoted by $G_c$, and a fully disconnected part,
denoted by $\overline{G_c}$, having an interface in between. The
interface is approximated with the solid-on-solid (SOS) model: having
a unique height $\zeta(u)$ at position $u$\cite{pottstm,long2d}. The
lattice contains $N=L^2$ points, from which $n(G_c)$ and
$n(\overline{G_c})$ are in $G_c$ and in $\overline{G_c}$,
respectively. Similarly, out of the $E=2L(L-1)$ edges, there are
$e(G_c)$ in the subgraph, $G_c$. The cost function in
Eq.(\ref{eq:kasfor1}) is given by:
\beqn
\phi&=&N+1-n(G_c)+\sum_{(u,v)\in G} (1+\kappa_0 \epsilon(u,v))/2\\ \nonumber
&=&N+1-n(G_c)-e(G_c)/2+\sum_{(u,v)\in G} \kappa_0 \epsilon(u,v)/2\;.
\eeqn
Now we use the microcanonical condition: $\sum_{all~bonds} \epsilon(u,v)=0$ and arrive to the cost function:
\be
4\phi(\zeta)=\kappa_0 \sum_{{u}}\left(\sum_{v<\zeta(u)}-\sum_{v>\zeta(u)}\right)\epsilon(u,v)\;,
\label{SOS2}
\ee
for a fixed value of $n(G_c)-e(G_c)/2$. This latter quantity is uniquely determined by the length of the interface,
which is {\it measured in the square lattice}.

Comparing the two expressions in Eqs.(\ref{SOS1}) and (\ref{SOS2}) we see that in the SOS approximation
the position of the interface is obtained by finding the extremal value of the same expression, however with
somewhat different constrains. In both cases the length of the interface is fixed, however in the RFIM
this length is measured in the medial lattice, whereas for the RBPM in the original lattice. In the following
we study both models numerically and see how well the SOS approximation and thus the above relation is valid
in finite systems.

\section{Evolution of interfaces with the strength of disorder}
\label{sec:Evol}
In the numerical study we use bimodal disorder, such that $\epsilon(u,v)=\pm 1$ with the same
probability.

\subsection{Random field Ising model}

\begin{figure}[h]
\begin{center}
\includegraphics[width=8.cm,angle=0]{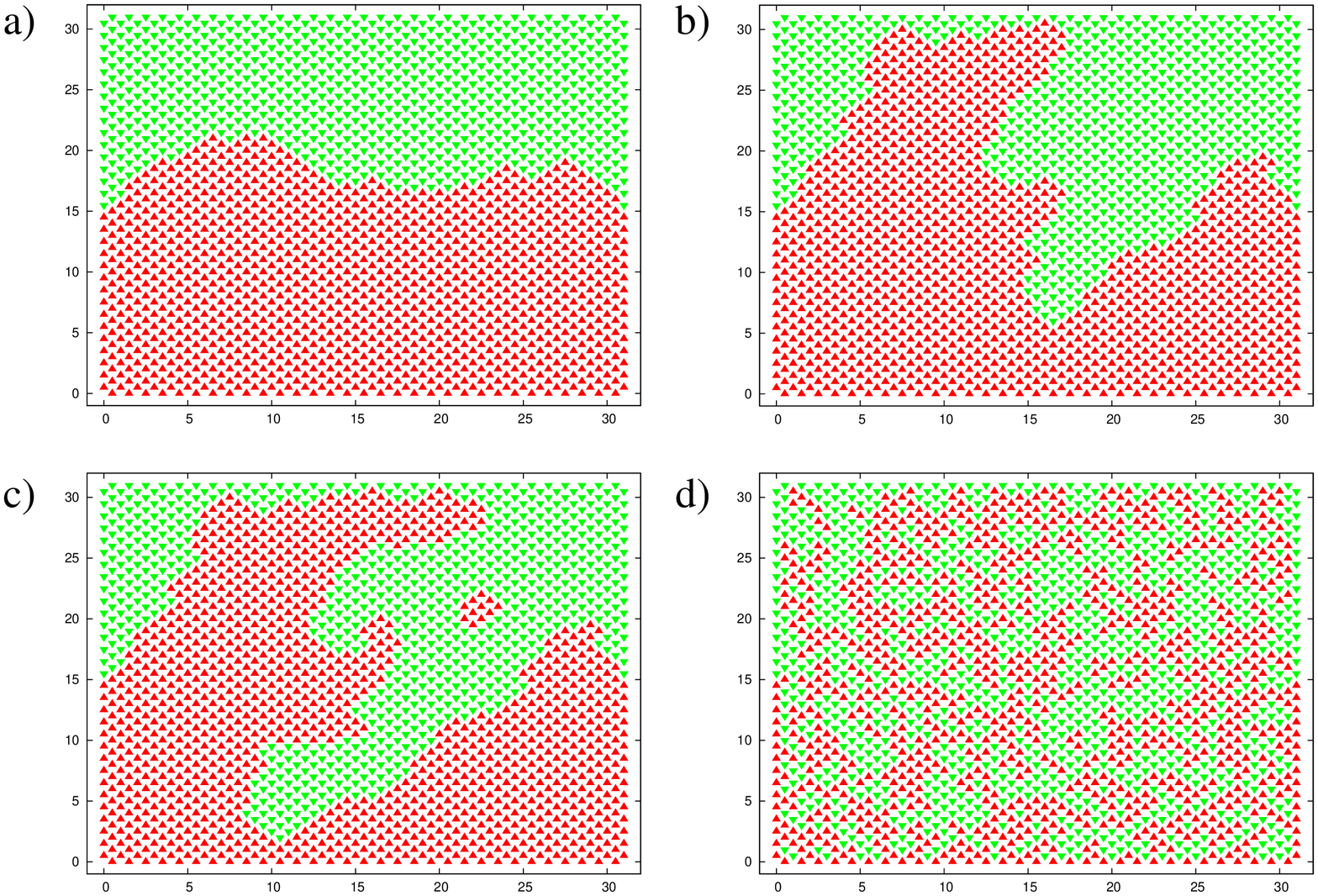}
\end{center}
\caption{Evolution of the ground state of the RFIM with increasing strength of disorder in a $32 \times 32$
lattice. Up and
down spins are represented by dark (red) and light (green) regions.
a) $h_0=0.1275$: Two clusters with an SOS interface. b) $h_0=0.2525$: Two clusters with an interface with overhangs.
c) $h_0=0.3275$: Internal
islands are formed within the big clusters. d) $h_0=4.0$: Several disjoint clusters are formed.
\label{fig2}
}   
\end{figure}

For RFIM the evolution of the ground state with the strength of disorder, $h_0$, for a given realization of the
random numbers, $\epsilon(u,v)$, is illustrated in Fig.\ref{fig2}. In the limit $h_0 \to 0^+$ there is a unique SOS
interface between two oppositely magnetized clusters. With increasing $h_0$ at $h_{0,1}$ the ground state will change. It
still consists of two large clusters, however the interface have some overhangs, thus it is no-longer of SOS
type. With further increased disorder, for $h_0 \ge h_{0,2}$, no longer two clusters exist, but some islands are
formed in the big clusters. Finally, for strong disorder, $h>h_{0,3}$, the ground state is represented by several disjoint
clusters and one can not identify an interface in the system\cite{percolation}.


\begin{figure}[h]
\begin{center}
\includegraphics[width=8.cm,angle=0]{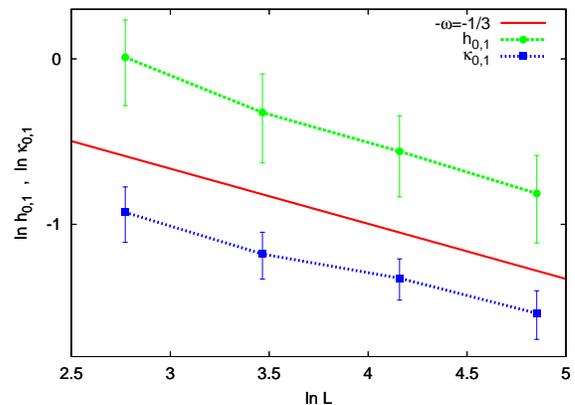}
\end{center}
\caption{ Average value of the critical disorder strength for the RFIM, $h_{0,1}$, and that for
the RBPM, $\kappa_{0,1}$, as a function of $L$ in log-log scale. The straight line indicating the
theoretical prediction has a slope $-\omega=-1/3$.
\label{fig3}
}
\end{figure}

We have studied the behavior of the first relevant disorder scale, $h_{0,1}$, at which the SOS picture is broken down.
Having several samples we have calculated its mean value, which is then repeated for different finite sizes, $L$.
$\overline{h}_{0,1}(L)$ is found to go to zero with $L$ as a power-low, $\overline{h}_{0,1}(L) \sim L^{-\omega}$, which is
illustrated in Fig.\ref{fig3}. The exponent is found to be: $\omega=0.36(5)$. Similarly, the second disorder scale, $h_{0,2}$,
is found to follow the same type of decay with $L$, with the same exponent, $\omega$, but with a different prefactor.


Our numerical results are in agreement with the exact result, that for any finite $h_0>0$ in the thermodynamic
limit there is no long range order in the $2d$ RFIM.

\subsection{Random bond Potts model}

\begin{figure}[h]
\begin{center}
\includegraphics[width=8.cm,angle=0]{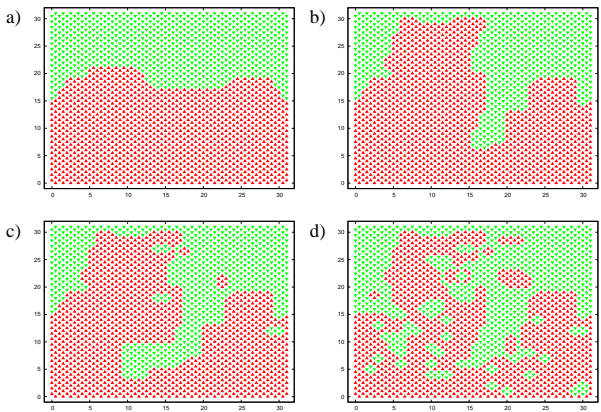}
\end{center}
\caption{Evolution of the optimal set of the RBPM with increasing
  strength of disorder in a $32 \times 32$ lattice having the same set
  of $\epsilon(u,v)$ as for the RFIM in Fig.\ref{fig2}. Connected
  graphs and isolated sites are represented by dark (red) and light
  (green) regions.  a) $\kappa_{0}=0.25$: Two clusters with an SOS
  interface. b) $\kappa_{0}=0.34$: Two clusters with an interface
  with overhangs. c) $\kappa_{0}=0.5$: Internal islands are formed
  within the big clusters. d) $\kappa_{0}=0.76$: Several disjoint
  clusters are formed.
\label{fig4}
}   
\end{figure}

We have repeated the previous analysis for the RBPM. For a given
random sample we have varied the strength of disorder, $\kappa_0$, and
studied the form of the optimal diagram. Having the same set of
$\epsilon(u,v)$ as for the RFIM in Fig.\ref{fig2} the results are
collected in Fig.\ref{fig4}. In the limit $\kappa_0 \to 0$ the
interface is plane, with a minimal length, $L_{int}=L$ (not shown in
Fig.\ref{fig4}). At a limiting value, $\kappa_{0,0}$, the interface
becomes an SOS interface with $L_{int}>L$. Further increasing
$\kappa_{0}$ the length of the interface is increasing, but it stays
SOS type. Afterwards, at $\kappa_{0,1}$ the interface will loose its
SOS character: overhangs and bubbles appear. Finally, for strong
disorder the optimal diagram is composed of disjoint subgraphs. We can
thus conclude that the evolution of the interfaces in the RBPM and in
the RFIM with the strength of disorder are qualitatively very similar.
There are, however, differences, for $\kappa_0 \le \kappa_0,0$, which
are due to the different orientations of the interface lines.

We have also studied the size dependence of the average value of $\kappa_{0,1}$, which is shown in Fig.\ref{fig3}.
Here also a power-low decay is found with the same exponent, $\omega$, as for the RFIM.


We can explain the value of the $\omega$ exponent, which is measured
both for the RFIM and for the RBPM, in the following way. Let us
concentrate now on the RFIM, for the RBPM similar reasoning works. At
$h_{0,1}$ the first overhang in the ground state of the RFIM
appears. The difference between the ground states for $h<h_{0,1}$ and
$h>h_{0,1}$ has a narrow shape. Its typical width is $w=O(1)$, since
the total length of the interface is increased by $\Delta
L_{int}=O(1)$, and its height is given by $L_{\perp}$, which is the
size of the transverse fluctuations of the SOS interface and given
by\cite{Gwa_Spohn}: $L_{\perp} \sim L^{2/3}$. Now we use the Imry-Ma
argument\cite{imryma} and compare the loss of energy due to the
increase of the interface: $E_{int} \sim w$, with that of the gain due
to disorder fluctuations: $E_{fl} \sim h_{0,1} \sqrt{w
  L_{\perp}}$. From the condition: $E_{int}=E_{fl}$ we obtain the
estimate $h_{0,1} \sim L^{-1/3}$, thus the exponent is given by
$\omega=1/3$ in agreement with the numerical results.

\section{Comparison of the structure of the graphs in the two models}
\label{sec:Comp}
\begin{figure}[h]
\begin{center}
\includegraphics[width=8.cm,angle=0]{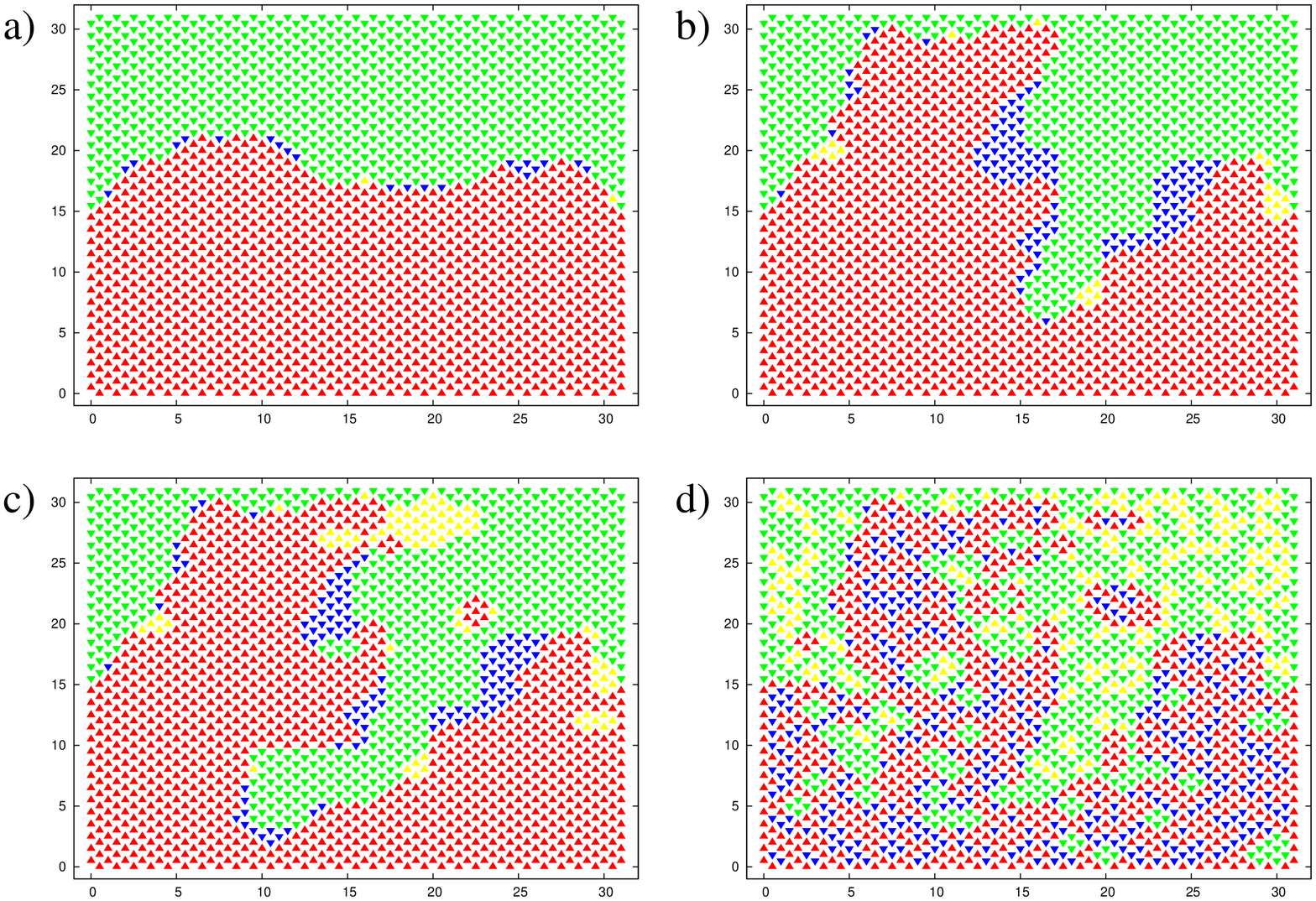}
\end{center}
\caption{Comparison of the ground state of the RFIM with the optimal
  set of the RBPM for the same pair of disorder parameters ($h_0$ and
  $\kappa_0$) as given in the panels of Figs.\ref{fig2} and
  \ref{fig4}, respectively.  Greyscale (gsc) 3 (red) points represent:
  up spins in the RFIM and the connected graph in the RBPM; gsc 2
  (green) points: down spins and isolated sites; gsc 4 (blue) points:
  down spins and connected graph; gsc 1 (yellow) points: up spins and
  isolated sites. The gsc 4 (blue) and gsc 1 (yellow) points are
  non-coherent sites.  The values of the discrepancies are
  $\delta=24/1984$, $135/1984$, $187/1984$ and $559/1984$, for panels
  $a)$, $b)$, $c)$ and $d)$, respectively.
\label{fig5}
}   
\end{figure}

The SOS mappings presented in Sec.\ref{sec:Model} assure that for weak
disorder the scaling behavior of the interfaces in the two models are
asymptotically equivalent. Here we study the question what happens for
finite systems and for not too weak disorder, which type of similarity
occurs between the ground states of the two problems. For this study
we consider exactly the same random samples for the two models, which
means that the set of random variables, $\epsilon(u,v)$, are the same
for each position, $(u,v)$, just the strength of disorder, $h_0$ and
$\kappa_0$, respectively, can be different. For a given sample we
calculate the ground state of the RFIM, as well as the optimal set of
the RBPM and compare them. In order to define a quantitative measure
of the difference between the two graphs we consider the medial
lattice and assign to each site, $(u,v)$, a variable denoted by
$D(u,v)$. If in the ground state of the RFIM $\sigma(u,v)=+1$
($\sigma(u,v)=-1$) and at the same time in the optimal set of the RBPM
the edge $(u,v)$ is occupied (non-occupied), then $D(u,v)=0$,
otherwise $D(u,v)=1$. In Fig.\ref{fig5} we compare the ground state
configurations in Fig.\ref{fig2} with the optimal sets in
Fig.\ref{fig4}.  One can see in this figure that close to the
boundaries $D(u,v)$ is typically zero (Greyscale (gsc) 3 (red) and gsc
2 (green) points) and in the interface region we have sites with
$D(u,v)=1$ (gsc 4 (blue) and gsc 1 (yellow) points).  The difference
between the two graphs is defined by the fraction of non-coherent
sites:
\be
\delta(h_0,\kappa_0)=\dfrac{1}{2L(L-1)}\sum_{(u,v)} D(u,v)\;,
\label{delta}
\ee
what we call as discrepancy. Here $2L(L-1)$ is the number of sites in the medial lattice with the given
boundary condition.

\begin{figure}[h]
\begin{center}
\includegraphics[width=8.cm,angle=0]{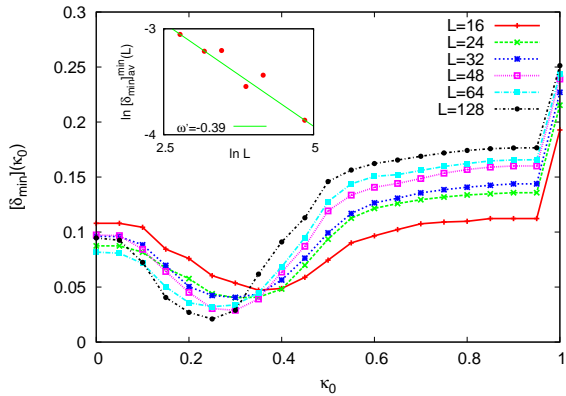}
\end{center}
\caption{Average value of the minimal discrepancy as a function of
the disorder parameter $\kappa_0$ for different finite systems. In the inset the minimal value of
$[\delta_{min}]_{\rm av}(\kappa_0)$ is plotted as a function of $L$ in a log-log scale. Here the straight
line has a slope of $\omega'=0.39$, to be compared with the theoretical prediction: $1/3$.
\label{fig6}
}   
\end{figure}

In the actual calculation we have fixed the value of the RBPM disorder
parameter, $\kappa_0$, and calculated the discrepancy for different
values of $h_0$. In this way we have measured the minimum value of the
discrepancy, $\delta_{min}(\kappa_0)$, and the corresponding value of
the RFIM parameter\cite{h0min}, $\tilde{h}_0$. The average value of
the minimal discrepancy, $[\delta_{min}]_{\rm av}(\kappa_0)$ versus
$\kappa_0$ is plotted in Fig.\ref{fig6} for different finite systems
from $L=16$ to $L=128$.


For large $\kappa_0$ the curves have a plateau, then with decreasing
$\kappa_0$ they start to decrease, pass over a minimum and afterwards
increase for small $\kappa_0$.  With increasing size the value at the
minimum $[\delta_{min}]_{\rm av}^{min}(L)$ is decreasing and in the
large-$L$ limit the minimum is expected to be shifted at the
origin. Indeed in the inset of Fig.\ref{fig6} we have plotted
$[\delta_{min}]_{\rm av}^{min}(L)$ as a function of $L$ and in a
log-log scale an asymptotically linear dependence is found, thus
$[\delta_{min}]_{\rm av}^{min}(L) \sim L^{-\omega'}$. Here the
exponent is $\omega'=0.39(7)$, which is compatible with $\omega'=1/3$,
the value of which follows from the following simple argument. For
weak disorder the interface in the RBPM is approximately flat (see the
reasoning in Sec.\ref{sec:Evol}) whereas in the RFIM it is rough
having a transverse fluctuation of size $L_{\perp} \sim L^{2/3}$. In
the interface region there are $L \times L_{\perp}$ sites,
consequently the average discrepancy behaves as $\sim L^{-1/3}$, in
agreement with the numerical results. We can thus conclude that in the
large-$L$ limit the behavior of the discrepancy with the strength of
disorder is the following. It starts from zero at $\kappa_0=0$, has an
approximately quadratic variation for small disorder and approaches a
plateau for stronger disorder. This behavior is in complete agreement
with the SOS mapping in Sec.\ref{sec:Model}).

We have also checked the actual form of the graphs in the two problems
for strong disorder. As illustrated in Fig.\ref{fig5} the two graphs
generally have quite similar structure, as far as the subgraphs, the
topology of the islands, etc.  are concerned.


\section{Discussion}
\label{sec:Disc}
In this paper we have considered two problems of lattice statistics in
which disorder has a strong and somewhat analogous effect as far as
the cooperative behavior of the models are considered. In both
problems geometrical interpretation of the state of the system is
used. In the first problem, which is the two-dimensional Ising model
we use clusters of parallel spins (geometrical clusters) to
characterize the ground state. Here random fields destroy the
ferromagnetic order at $T=0$ and the geometrical clusters have
homogeneous parts of
finite extent, $\ell_g(h_0)$. In the second problem, which is the
two-dimensional Potts model with a large number of states the optimal
diagram of the high-temperature series expansion is used as a
geometrical interpretation.  Here we consider the first-order
transition point of the pure system, which transition is softened to a
continuous one due to bond disorder. As a consequence at the
transition point there is no phase-coexistence and the homogeneous parts
of the optimal diagram have a finite extent of $\ell_d(\kappa_0)$.

In this paper we have considered finite samples of linear length, $L$,
by varying the strength of disorder, $h_0$ and $\kappa_0$ in the two
problems, respectively. Having fixed spin boundary conditions for
$\ell_g(h_0)>L$ ($\ell_d(\kappa_0)>L$) there are basically two phases
(two elementary diagrams) in the ground state (optimal diagram) of the
RFIM (RBPM), which are separated by an interface. For large scales and
in the SOS approximation the interface Hamiltonians of the two
problems have similar forms.

Here we have studied the validity of the interface mapping in finite
systems, by putting the Potts model on the square lattice and the
Ising model on the medial lattice. In this way the bonds of the Potts
model and the fields of the Ising model have the same location. Using
the same disordered samples we have studied the evolution of the
interfaces in the two models as the strength of disorder is gradually
increased. We have seen a similar trend of the evolution in the two
models, however, in finite lattices we have observed also
differences. These differences are basically due to the fact, that the
interfaces have different orientations in the two models: in the RBPM
it is in the (1,0) direction, whereas in the RFIM it has an (1,1)
orientation. In the SOS approximation the interfaces are obtained by
minimizing the same cost function, however with different constrains
in the two problems.

We have also compared the cluster structure of the two problems and
have obtained the following conclusion.  The relative difference
between the two geometrical objects as quantified by the discrepancy
in Eq.(\ref{delta}) tends to zero if i) the size of the system goes to
infinity and ii) the strength of disorder goes to zero. The
finite-size corrections are found to be in power law form, which
vanish as $L^{-\omega}$ with $\omega=1/3$. Consequently relatively
large finite samples are needed to see the asymptotic behavior.

\begin{acknowledgments}
This work has been supported by the Hungarian National Research Fund under grant No OTKA
K62588, K75324 and K77629. F.I. is indebted to the Institut N\'eel-MCBT for hospitality
during the final stages of the work. M.K. thanks the Minist\`ere Fran\c{c}ais des
Affaires \'Etrang\`eres for a research grant.
\end{acknowledgments}


\begin{thebibliography}{99}
\bibitem{percolation}
	This point should be related to the percolation transition of the geometrical clusters\cite{seppala,kornyei}.
	For $h<h_{0,3}$ there is an infinite geometrical cluster in the system, whereas for stronger random
	fields there is no spanning cluster.
\bibitem{h0min}
        For a given value of $\kappa_0$ there is a finite interval of $h_0$ which corresponds
        to the minimal discrepancy condition. Here we use the smallest such value as $\tilde{h}_0$.
\end{thebibliography}
\end{document}